\documentclass[a4paper,11pt]{article}
\usepackage{amsfonts}
\usepackage{amsmath}

\input{tcilatex}

\begin{document}

\title{Recent CMB observations enable to find the total gravitational energy
of a mass}
\author{Dimitar Valev \\
\textit{Stara Zagora Department, Solar-Terrestrial Influences Laboratory,}\\
\textit{Bulgarian Academy of Sciences, 6000 Stara Zagora, Bulgaria}}
\maketitle

\begin{abstract}
The astronomical observations indicate that the universe expands with
acceleration and it has a finite particle horizon. The recent \textit{CMB}
observations confirm the universe is homogeneous, isotropic and
asymptotically flat. The total gravitational energy of a body having mass $m$
is the gravitational potential energy originating from the gravitational
interaction of the body with all masses of the observable universe, i.e.
within the particle horizon. The flat geometry of the universe enables to
determine the total gravitational energy of the mass $m$ within the
framework of the Newtonian gravity in Euclidean space. By this approach, it
has been found the modulus of the total gravitational energy of a body is
close to its rest energy $E=mc^{2}$, which is a remarkable result. Besides,
the smoothed gravitational potential in an arbitrary point of the observable
universe appears close to $-c^{2}$, where $c$ is the speed of the light.

\end{abstract}

\section{Introduction}

In Big Bang cosmology, the observable universe consists of the galaxies and
other matter that we can in principle observe from Earth in the present day,
because light (or other signals) from those objects has had time to reach us
since the beginning of the cosmological expansion. The total gravitational
energy of a body having mass $m$ is the gravitational energy of the mass $m$%
, originating from the gravitational interaction of the body with all masses
in the universe. This quantity obtains limited value if the universe (or
particle horizon) is finite. Besides, to determine the total gravitational
energy of a mass $m$, the geometry and density of the universe need to be
known.

The problem for the total average density of the universe $\overline{\rho }$%
\ acquires significance when it has been shown that the General Relativity
allows to reveal the large-scale structure and evolution of the universe by
simple cosmological models \cite{Friedman 1922, Lemaitre 1927, Einstein 1932}%
. Crucial for the geometry of the universe appears dimensionless total
matter density $\Omega =\overline{\rho }/\rho _{c}$, where $\rho _{c}$\ is
the critical density of the universe. The most trustworthy total density $%
\Omega $ has been determined by measurements of the dependence of the
anisotropy of the Cosmic Microwave Background (\textit{CMB}) upon the
angular scale. The recent results show that $\Omega =1\pm \Delta \Omega $,
where the error $\Delta \Omega $ decreases from 0.10 \cite{de Bernardis
2000, Balbi 2000} to 0.02 \cite{Spergel 2003}. The fact that $\Omega $ is so
close to a unit is not accidental since only at $\Omega =1$ the geometry of
the universe is flat (Euclidean) and the flat universe was predicted from
the inflationary theory \cite{Guth1981}. The total density $\Omega $
includes densities of baryon matter $\Omega _{b}\approx 0.05$, cold dark
matter $\Omega _{c}\approx 0.25$ \cite{Peacock 2001} and dark energy $\Omega
_{\Lambda }\approx 0.70$ producing an accelerating expansion of the universe 
\cite{Riess 1998, Perlmutter 1999}.

The found negligible \textit{CMB} anisotropy $\frac{\delta T}{T}\sim 10^{-5}$
indicates that the early universe was very homogeneous and isotropic \cite%
{Bennett 1996}. Three-dimensional maps of the distribution of galaxies
corroborate homogeneous and isotropic universe on large scales greater than
100 $Mps$ \cite{Shectman 1996, Stoughton 2002}. In the present paper, the
recent \textit{CMB} observations are used to determine the total
gravitational energy of a body having mass $m$ placed in an arbitrary
location, far away from strong local gravitational fields. Such fields
appear close to neutron stars, black holes, nuclei of galaxies and quasars.

\section{Determination of the total gravitational energy of a body}

Finite Hubble time $H^{-1}$ (age of the universe) and finite speed of light $%
c$ set a \textit{finite} particle horizon beyond which no material signals
reaches the observer. As a result, a body having mass $m$ interacts
gravitationally with all masses $m_{i}$ at distances $r_{i}<R$, where $R\sim
c/H$ is the Hubble distance and $H=H_{0}h\approx $ 70 $km$ $s^{-1}$ $%
Mps^{-1} $ is the Hubble expansion rate \cite{Mould 2000}. All these masses
form the causally connected universe. Total gravitational energy of a body
having mass $m$ is the finite gravitational energy of the mass $m$,
originating from the gravitational interaction of the body with all masses
within the particle horizon.

The customary approach used for such cosmological problems is in the
framework of the General Relativity, since at cosmological distances the
space curvature should be taken into considerations. But, on account of the
total density $\Omega =1$, the global geometry of the universe appears flat
and the space curvature is zero. This enables to apply Newtonian gravity in
Euclidean space for solution of this cosmological problem.

Thus, the problem of the total gravitational energy of a mass $m$ transforms
into the classical problem of the gravitational potential in the centre of a
homogeneous isotropic sphere having a finite radius $R\sim c/H$ and density $%
\overline{\rho }=\Omega \rho _{c}$. Therefore, the total gravitational
energy $U$ of a mass $m$ in the homogeneous and isotropic universe, far away
from strong local gravitational fields, would be expressed by equation:

\begin{equation}
U=-Gm\tsum\limits_{i}\frac{m_{i}}{r_{i}}=-4\pi mG\overline{\rho }%
\tint\limits_{0}^{R}rdr=-2\pi mG\overline{\rho }R^{2}  \label{Eqn1}
\end{equation}

where 0 is an \textit{arbitrary} location of an observer and $R\sim c/H$ is
his particle horizon, i.e. his observed `radius' of the universe. The
integration of (\ref{Eqn1}) is made in Euclidean space.

This approach has been used in \cite{Woodward 1975} for another problem,
namely the estimation of the graviton mass. According the authors, Newtonian
gravity is acceptable for the calculation of the total gravitational energy
even in the case of $\Omega \ll 1$. Still more the applied approach would be
adequate in the case of $\Omega \approx 1$ \cite{Valev 2009}.

The critical density of the universe $\rho _{c}$ determines \cite{Peebles
1971} from equation:

\begin{equation}
\rho _{c}=\frac{3H^{2}}{8\pi G}  \label{Eqn2}
\end{equation}

In view of $\overline{\rho }=\Omega \rho _{c}$ and (\ref{Eqn2}), the
equation (\ref{Eqn1}) transforms into:

\begin{equation}
U=-\frac{3}{4}\Omega mR^{2}H^{2}  \label{Eqn3}
\end{equation}

In consideration of $R\sim c/H$, we obtain:

\begin{equation}
U\approx -\frac{3}{4}\Omega mc^{2}=-\frac{3}{4}mc^{2}  \label{Eqn4}
\end{equation}

The equation (\ref{Eqn4}) shows that the modulus of the gravitational energy
of a body, originating from the gravitational interaction of the body with
all masses within the particle horizon, is approximately (with accuracy to a
factor 3/4) equal to its \textit{rest energy} $E=mc^{2}$. Thus, the rest
energy of an arbitrary mass $m$ is approximately balanced with its total
gravitational energy. In result the total energy of an arbitrary mass $m$,
including its total gravitational energy, is close to zero.

The factor 3/4 in (\ref{Eqn4}) most likely arises as a result of the use of
the approximation $R\sim c/H$ in equation (\ref{Eqn3}). This approximation
is valid with accuracy to the coefficient $k\sim 1$ depending on the
specific cosmological model of the expansion, i.e. $R=kc/H$. Clearly, for $k=%
\sqrt{4/3}\approx $ 1.155 the equation (\ref{Eqn4}) of the total
gravitational energy of a mass $m$ will be replaced from equation:

\begin{equation}
U=-\frac{3}{4}k^{2}\Omega mc^{2}=-\Omega mc^{2}=-mc^{2}  \label{Eqn5}
\end{equation}

According to the definition, the total gravitational energy $U$ of the mass $%
m$ is equal to the work, which does the gravity originating by all masses in
the causally connected universe for a removal of the mass $m$ from its
current location to the infinity. Therefore, the rest energy $E=mc^{2}$ of a
mass $m$ is close to the gravitational energy, which would be released if
the mass were moved from the infinity to its current location.

The \textit{smoothed} gravitational potential $\varphi $ in an arbitrary
point of the homogeneous and isotropic universe, far away from strong local
gravitational fields, follows from (\ref{Eqn4}):

\begin{equation}
\varphi =\frac{U}{m}\approx -\frac{3}{4}\Omega c^{2}=-\frac{3}{4}c^{2}
\label{Eqn6}
\end{equation}

The equation (\ref{Eqn6}) shows that the smoothed gravitational potential in
an arbitrary point of the observable universe appears close to $-c^{2}$,
where $c$ is the speed of the light. Since the observable universe appears
equipotential 3-dimensional sphere, no additional (cosmological)
gravitational force acts on the masses.

According (\ref{Eqn6}), the smoothed gravitational potential $\varphi $ in
an arbitrary point of the homogeneous and isotropic universe depends
linearly from the density of the universe $\Omega $. Clearly, only in a case
of $\Omega \sim 1$, the smoothed gravitational potential is $\varphi \sim
-c^{2}$. If the universe was consisted of baryonic matter only, than total
density $\Omega \approx $ 0.05 and $\left\vert \varphi \right\vert \ll c^{2}$%
, but the high densities of the cold dark matter and dark energy increase
the density to $\Omega =1$. In result, the universe appears flat and the
modulus of the gravitational energy of a body having mass $m$ is close to
its rest energy.

\section{Conclusions}

The astronomical observations indicate that the accelerating universe has a
finite particle horizon. The recent \textit{CMB} observations confirm the
universe is homogeneous and isotropic on large scales and the geometry is
asymptotically flat. The flat geometry of the universe enables to determine
the total gravitational energy of the mass $m$ within the framework of the
Newtonian gravity in Euclidean space. Thus, the problem of the total
gravitational energy of a mass $m$ transforms into the classical problem of
the gravitational potential in the centre of a homogeneous isotropic sphere
having a finite radius $R\sim c/H$.

By this approach, it has been found that the modulus of the gravitational
energy of a body, originating from the gravitational interaction of the body
with all masses within the particle horizon, is close to its rest energy $%
E=mc^{2}$. Thus, the rest energy of an arbitrary mass $m$ is approximately
balanced with its total gravitational energy. In result the total energy of
an arbitrary mass $m$, including its total gravitational energy, is close to
zero. Besides, the smoothed gravitational potential in an arbitrary point of
universe is close to $-c^{2}$. Finally, it has been shown that these
evaluations are valid only in a case of $\Omega \sim 1$, i.e. in a flat
universe.


\bigskip

\begin{thebibliography}{10}
\bibitem[1]{Friedman 1922} Friedman A., Z. Physik, 10, 1922, 377.

\bibitem[2]{Lemaitre 1927} Lemaitre G., Ann. Soc. Sci. Brux., 47A, 1927, 49.

\bibitem[3]{Einstein 1932} Einstein A., W. de Sitter, Proc. Nat. Acad. Sci.
USA, 18, 1932, 213.

\bibitem[4]{de Bernardis 2000} de Bernardis P. et al., Nat., 404, 2000, 955.

\bibitem[5]{Balbi 2000} Balbi A. et al., ApJ., 545, 2000, L1.

\bibitem[6]{Spergel 2003} Spergel D. N. et al., ApJS, 148, 2003, 175.

\bibitem[7]{Guth1981} Guth A. H., Phys. Rev. D, 23, 1981, 347.

\bibitem[8]{Peacock 2001} Peacock J. A. et al., Nat., 410, 2001, 169.

\bibitem[9]{Riess 1998} Riess A. G. et al., AJ., 116, 1998, 1009.

\bibitem[10]{Perlmutter 1999} Perlmutter S. et al., ApJ., 517, 1999, 565.

\bibitem[11]{Bennett 1996} Bennett C. L. et al., ApJ., 464, 1996, L1.

\bibitem[12]{Shectman 1996} Shectman S. A. et al., ApJ., 470, 1996, 172.

\bibitem[13]{Stoughton 2002} Stoughton C. et al., AJ., 123, 2002, 485.

\bibitem[14]{Mould 2000} Mould J. R. et al., ApJ., 529, 2000, 786.

\bibitem[15]{Woodward 1975} Woodward J. F. et al., Phys. Rev. D, 11, 1975,
1371.

\bibitem[16]{Valev 2009} Valev D., Compt. rend. Acad. bulg. Sci., Special
Issue: Fundamental Space Research, 2009, 233; http://arxiv.org/abs/0909.2726.

\bibitem[17]{Peebles 1971} Peebles P. J. E., 1971, Physical Cosmology,
Princeton Univ. Press, Princeton, NJ.
\end{thebibliography}
\end{document}